\documentstyle[psfig,amstex,amssymb]{mn}

\def\kmsmpc{\rm{kms^{-1}Mpc^{-1}}}
\def\etal{\rm{et al.}}

\def\refer{\par \noindent \hangindent=3pc \hangafter=1}
\def\oml{\Omega_{\Lambda}}
\def\om{\Omega_{\rm{m}}}
\def\zeff{z_{{\rm eff}}}
\def\zqso{z_{{\rm qso}}}
\def\zgal{z_{{\rm gal}}}

\begin{document}

\title[No Periodicities in 2dF Redshift Survey Data] 
{No Periodicities in 2dF Redshift Survey Data}

\author[E. Hawkins, S.J. Maddox and M.R. Merrifield]
{E. Hawkins\footnotemark, S.J. Maddox and M.R. Merrifield\\
School of Physics \& Astronomy, University of Nottingham, Nottingham
NG7 2RD, UK}

\maketitle

\date{}

\begin{abstract}
We have used the publicly available data from the 2dF Galaxy Redshift
Survey and the 2dF QSO Redshift Survey to test the hypothesis that
there is a periodicity in the redshift distribution of quasi-stellar
objects (QSOs) found projected close to foreground galaxies.  These
data provide by far the largest and most homogeneous sample for such a
study, yielding 1647 QSO--galaxy pairs.  There is no evidence for a
periodicity at the predicted frequency in $\log(1+z)$, or at any other
frequency.
\end{abstract}


\section{Introduction} 
\footnotetext{E-mail: ppxeh@@nottingham.ac.uk} 

Claims of periodicities or regularities in redshift distributions of
various astronomical objects have been made for many years
(e.g. Burbidge \& Burbidge 1967; Broadhurst \etal, 1990; Karlsson,
1990; Burbidge \& Napier, 2001).  This effect, if real, has
far-reaching implications for the interpretation of redshift as a
cosmological phenomenon, and, indeed, for the nature of objects like
quasi-stellar objects (QSOs) that appear to display the periodicities.

One particularly intriguing effect has been explored by Arp et al.\
(1990) and Karlsson (1990) and extended to a larger sample by Burbidge
\& Napier (2001).  It involves the apparent strong periodicity in
$\log(1+\zqso)$ for a sample of QSO redshifts, $\zqso$, where the QSO
appears projected close to a ``foreground'' galaxy at lower redshift.
If confirmed, such an effect would be impossible to explain in
conventional cosmological terms: it would either require that the QSOs
be physically associated with the galaxies in an as-yet unexplained
fashion, or that the QSO light passing the galaxy is somehow
influenced to quantize its redshift.

The criticism usually levelled at this kind of study is that the
samples of redshifts have tended to be rather small and selected in a
heterogeneous manner, which makes it hard to assess their
significance.  The more cynical critics also point out that the
results tend to come from a relatively small group of astronomers who
have a strong prejudice in favour of detecting such unconventional
phenomena.  This small group of astronomers, not unreasonably,
responds by pointing out that adherents to the conventional
cosmological paradigm have at least as strong a prejudice towards
denying such results.

In an attempt to circumvent these problems, Bill Napier contacted the
authors of this paper.  The availability of the data from the 2dF
Galaxy redshift Survey (2dFGRS) and the 2dF QSO Redshift Survey (2QZ)
means that for the first time there exists a large homogeneous sample
of data to carry out this kind of study.  Furthermore, Napier
recognized the importance of the study being carried out independent
from any of the researchers with vested interests one way or the
other.  He therefore gave clear instructions as to what analysis
should be performed and what periodic effect should be seen if the
phenomenon is real, but chose to take no part in the subsequent
analysis.  We have attempted to carry out this analysis without
prejudice.  Indeed, we would have been happy with either outcome: if
the periodicity were detected, then there would be some fascinating new
astrophysics for us to explore; if it were not detected, then we would
have the reassurance that our existing work on redshift surveys, etc,
has not been based on false premises.

The remainder of this paper is laid out as follows.
Section~\ref{s:napier} presents Napier's prediction as to what signal
we should expect to see if the data are analysed appropriately.
Section~\ref{s:data} describes the data set, and Section~\ref{s:meth}
presents the manner in which it has been analysed.  The results are
described in Section~\ref{s:res}.

\section{Napier's Prediction}
\label{s:napier}

In briefing us, Bill Napier stated that a strong periodicity had been
found in the redshifts of QSOs projected within $30\,{\rm armin}$ of
the centres of nearby galaxies (either in the Virgo Cluster or bright
galaxies in the Shapley Ames Catalog), corresponding to a physical
scale of $\sim 200\,{\rm kpc}$ at these galaxies' distances.  He
therefore suggested that we look at all galaxies in the 2dFGRS, use
their redshifts to estimate their distance (adopting a Hubble constant
of $60\kmsmpc$), and find all QSOs from the 2QZ survey projected
within a circle whose radius corresponds to $200\,{\rm kpc}$ at the
galaxy's distance.  Then, after transforming the QSOs' redshift to the
reference frame of the galaxies that they lie behind, we should expect
to find a strong periodic signal in $\log(1 + z)$ at a period $P \sim
0.09$.

\section{The 2dF Data}
\label{s:data}

\subsection{The parent catalogues}

For this study, we use the large databases provided by two surveys
carried out using the 2dF multi-object spectrograph on the
Anglo-Australian Telescope (Lewis \etal\ 2002).  For the galaxies, we
use the publicly available 100k data release from the 2dF Galaxy
Redshift Survey (Colless \etal~2001) and for the QSOs we take the
publicly available data from the 10k 2dF QSO Redshift Survey (Croom
\etal~2001). These two surveys used shared observations to measure the
redshifts of well-defined samples of galaxies and QSOs in a common
region of the sky, making them ideally suited to this analysis. 

To ensure the reliability of our sample we only use galaxies from the
2dFGRS with the two highest quality flags, $Q \geq 4$, in order to
have confidence in the derived redshifts, and consider only those
galaxies in the range $0.01 < z < 0.3$.  For the 2QZ sample we only
use QSOs with the highest quality flag in the database, which implies
a clear spectral identification of the object as a QSO.  This quality
control leaves a total of 67291 galaxies and 10410 QSOs in the
samples.

The survey strategies and the limitations of the 2dF instrument mean
that the selection of objects in the galaxy and QSO surveys are not
entirely independent.  For example, the diameters of the individual
fibers mean that very close pairs of objects are sometimes missed.
However, none of these geometric selection criteria depend on the
redshifts of the galaxies or QSOs, so although the sample may be
somewhat incomplete, its redshift distribution will not be biased in
any way by the selection process.

\subsection{Pair selection}
As instructed, we have inter-compared these datasets to find all
QSO--galaxy pairs with an angular separation corresponding to less
than 200~kpc at the distance of the galaxy.  For this calculation, we
adopt the fashionable $\Lambda$ cosmology, with parameters $\om = 0.3,
\oml = 0.7$ and $H_0 = 60\kmsmpc$.  However, the relatively low
redshifts of the foreground galaxies means that the choice of
cosmology makes essentially no difference to the sample selection.  In
a number of cases there is more than one galaxy within the 200 kpc
projected distance limit of the QSO; for these objects we take the
closest galaxy in projected distance to make up the pair.  In a few
cases, the same galaxy may be used for more than one QSO.  This
procedure yields a total of 1647 QSO--galaxy pairs.

\begin{figure}
\psfig{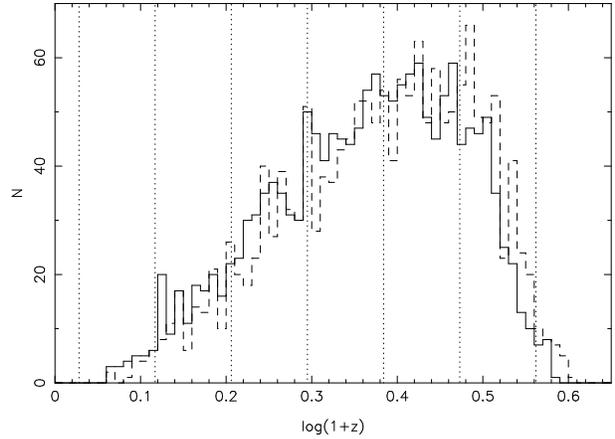}
\caption[]{Histograms of $\log(1+z)$ for for the 2dF close pairs using
$\zeff$ (solid line) and $\zqso$ (dashed line). The vertical dotted
lines are the locations of peaks in the redshift distribution as
predicted by Burbidge \& Napier (2001).}
\label{f:hists}
\end{figure}

The predicted periodicity lies in $\log(1 + \zeff)$, where $\zeff$ is
the redshift of the QSO measured relative to the nearby galaxy, so we
define 
\begin{equation}
1 + \zeff = (1 + \zqso)/(1 + \zgal),
\end{equation}
where $\zqso$ and $\zgal$ are the corresponding heliocentric
measurements for the QSO and galaxy.  Figure~\ref{f:hists} shows the
distribution of $\zeff$ and $\zqso$ for the 1647 pairs, with the
locations of the predicted periodic peaks indicated.  No periodicity
leaps off the page, but since the effect is likely to be quite subtle,
one would not necessarily expect to be able to pick it out from the
raw data, so it is important to carry out a rigorous statistical
analysis.

\section{Analysis}
\label{s:meth}

\subsection{The Power Spectrum}
We wish to measure the power spectrum for a set of $N$ measurements of
some quantity $x_i$ [in the current case, the value of $\log(1 +
\zeff)$ for different QSO--galaxy pairs].  Following the conventions
of Burbidge \& Napier (2001), we define the power $I$ at period $P$
via the formulae
\begin{equation}
I(P) = 2 R^2 / \sum^N_{i=1} w_i^2
\end{equation}
where
\begin{equation}
R^2 = S^2 + C^2
\end{equation}
with 
\begin{equation}
S = \sum^N_{i=1} w_i \sin(2 \pi x_i/P),\hspace*{0.4cm}C =
\sum^N_{i=1} w_i \cos(2 \pi x_i/P).
\label{eq:cssum}
\end{equation}
The quantity $w_i$ is a weighting function to apodize any ill effects
from the window function (see below); in the Burbidge \& Napier (2001)
analysis, $w_i \equiv 1$.  With this definition of the power spectrum,
an infinite uniform random distribution of values $x_i$ would yield
$I \equiv 2$.

Error bars on $I(P)$ can be estimated using the ``jackknife''
technique (Efron 1979) of drawing all possible samples of $N-1$ values
from the $N$ data points (without replacement), repeating the power
spectral analysis on these resamplings, and calculating the standard
deviation in the derived values of $I$ at different periods $P$,
$\sigma_J(P)$.  The best estimator for the standard error in the value
of $I$ is then just $\sqrt{N - 1}\sigma_J$.

\subsection{The Window Function}
The above analysis works perfectly for detecting periodic signals in
data sets of infinite extent.  However, in practice, such analyses are
based on data sets that are finite in extent.  In particular, the
redshift distribution of QSOs has a cut-off at low redshifts due to
the small volume sampled, and one at high redshift due to the colour
selection by which QSOs are found.  In addition, there may well be
variations in average numbers due to evolution in the quasar
population with redshift.  Thus, the idealized infinite data series is
truncated to a finite series by a ``window function'', which varies
from a value of unity where no data are missed to zero outside the
range sampled.

\begin{figure}
\psfig{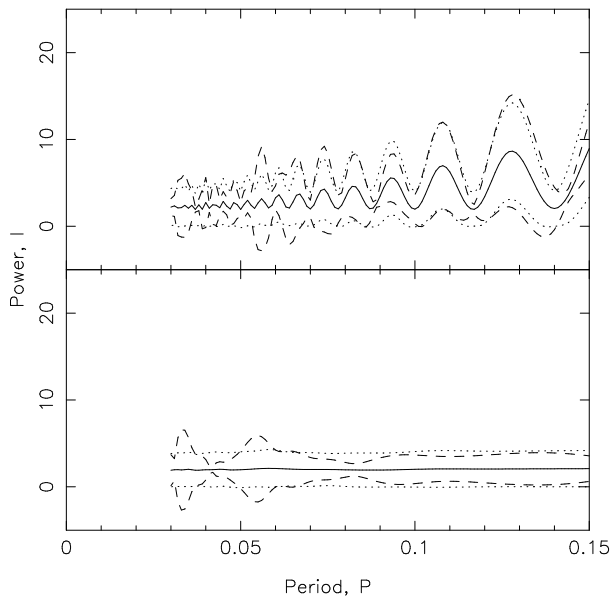}
\caption[]{Simulation of the power spectrum of a truncated uniform
distribution.  Each simulation drew 1000 measurements, uniformly
distributed between $0$ and $0.7$.  The solid line shows the mean of
1000 realizations, while the dotted lines shows their standard
deviation.  The dashed lines are the errors as derived by applying the
jackknife estimator to a single simulation.  The upper panel shows the
raw power spectrum, and the lower panel shows the power spectrum
derived with the data weighted using a Hann function.}
\label{f:flat}
\end{figure}

This truncation can introduce strong spurious features into the power
spectrum.  Its impact depends quite sensitively on how sharply the
cut-off occurs.  As an extreme example, the upper panel of
Fig.~\ref{f:flat} shows a simulation of a uniform random distribution
truncated sharply at $0$ and $0.7$, corresponding crudely to the
distribution in $\log(1+z)$ of the data presented in Burbidge \&
Napier (2001).  As can be seen from this figure, the window function
introduces many seemingly-periodic features, so that the resultant
spectrum differs greatly from the $I \equiv 2$ that one would expect
for an infinite uniform distribution.  This figure also shows that the
jackknife error analysis does a good job in determining the true
root-mean-square uncertainty in the power spectrum.

In practice, the window function is unlikely to cut off this sharply,
so the effects will be rather smaller than this extreme case.
Nonetheless, particularly for relatively small periodic signals, it is
vital that the effects of the window be taken into account [a point,
indeed, noted by Burbidge \& Napier (2001) in their analysis of
Karlsson's (1990) data, where one peak is disregarded as just such an
artifact].  However, with relatively small samples taken from
heterogeneous data sets, it is very difficult to formally quantify the
selection function that specifies the shape of the window, so a
rigorous analysis is difficult to implement.  

Fortunately, without knowing the exact nature of the window function
of the sample, one can manipulate the data in order to specify ones
own more optimal window -- a procedure that statisticians whimsically
refer to as ``carpentry.''  This process involves reducing the
weighting of data close to the ends of the range observed, thereby
smoothing off the sharp edges of the window, or ``apodizing'' the
function.  We achieve this apodization by using the Hann function as a
weighting,
\begin{equation}
w_i = {1 \over 2} \left[1 - \cos\left({2\pi x_i \over L}\right)\right]
\end{equation}
in equation \ref{eq:cssum}.  Here, $L$ is chosen to cover the range
over which the data are selected, so that the weighting goes smoothly
from unity in the middle of the range to zero at $x_i = 0$ and $x_i =
L$.  As the lower panel of Figure~\ref{f:flat} shows, even for the
most extreme possibility of an intrinsically instantaneous cut-off in
the window function, this procedure effectively restores the power
spectrum to close to the expected value of $I(P) \equiv 2$, without
compromising the errors derived using the jackknife analysis.

\begin{figure}
\psfig{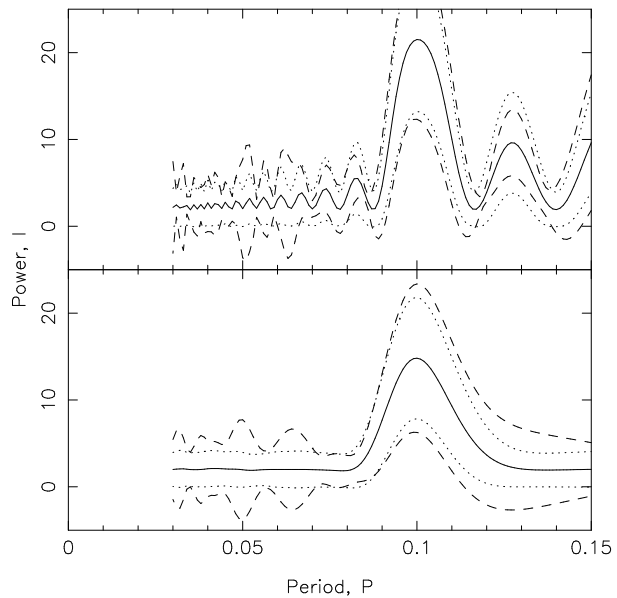}
\caption[]{As for Figure~\ref{f:flat}, but with a small periodic
signal added to the noise.}
\label{f:per}
\end{figure}

One concern with such a process is that it could erase real periodic
signals as well as the spurious artifacts.  To address this issue, we
repeated the above simulation with a small periodic addition to the
uniform distribution.  As Figure~\ref{f:per} shows, the apodization of
the data does not erase the periodic signal, and, once again, the
jackknife errors provide a good estimate for the true uncertainty.

\section{Results}
\label{s:res}

\begin{figure}
\psfig{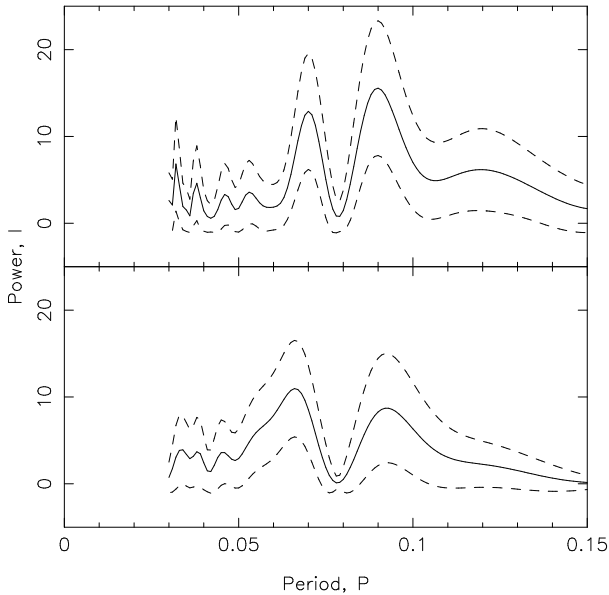}
\caption[]{Power spectrum analysis of the data from Karlsson (1990).
The upper panel shows the raw power spectrum [very similar to that in
Fig.~1 of Burbidge \& Napier (2001)]; the lower panel shows and the
spectrum after apodization with a Hann function.  Jackknife RMS
uncertainties are indicated as dashed lines.}
\label{f:karl}
\end{figure}
\subsection{The Karlsson data set}
As an initial test of the code developed for this study, we have
reanalysed the 116 QSO redshifts from Karlsson (1990) to make sure
that we can reproduce the results derived in Burbidge \& Napier
(2001).  As Figure~\ref{f:karl} shows, the unweighted spectral
analysis reveals the peak in the spectrum at $P \sim 0.09$ that
Burbidge \& Napier (2001) detected, as well as the peak at $P \sim
0.07$ that they attributed to the window function.  As expected for
this latter artifact, when the data are apodized, its strength is
reduced to an insignificant $\sim 1.5\sigma$ above the noise value
of $I = 2$.  However, the stronger ``real'' signal is even more
dramatically reduced to a significance of only $\sim 1\sigma$.  This
analysis would indicate that the peak at $P \sim 0.09$ may well be
compromised by the window function in this data set.

\begin{figure}
\psfig{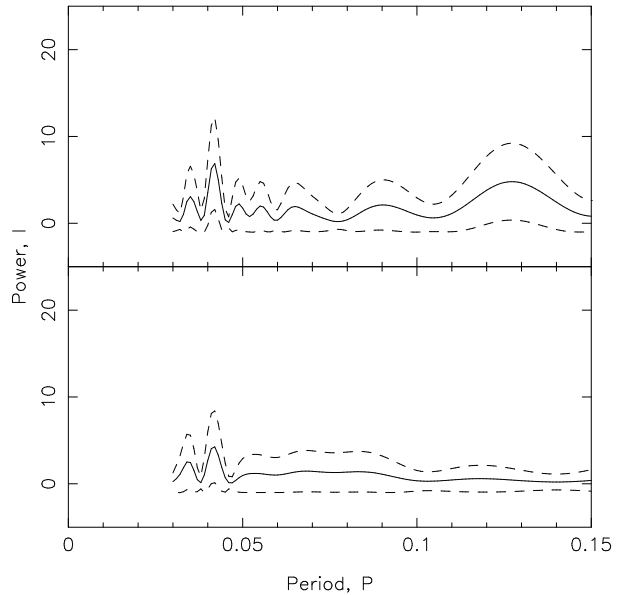}
\caption[]{Power spectrum analysis of the QSO--galaxy pairs in the 2dF
survey data.  The upper panel shows the raw power spectrum, and the
lower panel shows and the spectrum after apodization with a Hann
function.  Jackknife RMS error estimates are shown as dashed lines.}
\label{f:twodf}
\end{figure}

\subsection{The 2dF data set}
Figure~\ref{f:twodf} shows the same analysis applied to the sample of
1647 QSO--galaxy pairs drawn from the 2dF surveys, as described in
Section~\ref{s:data}.  Here, the raw and apodized power spectra are
quite similar -- the apodization's lack of major impact presumably
reflects the overall smooth distribution in Fig.~\ref{f:hists}, which
is already quite close to optimal in shape.  In any case, it is
apparent that there is no significant periodicity in the data at $P
\sim 0.09$, or, indeed, at any other frequency.  An analysis of
the QSOs' heliocentric redshifts revealed a similar absence of
significant periodicities.  Given that there are almost eight times as
many data points in this sample as in the previous analysis by
Burbidge \& Napier (2001), we must conclude that the previous
detection of a periodic signal arose from the combination of noise and
the effects of the window function.

\section*{Acknowledgements}
We are most grateful to Bill Napier for suggesting that we undertake
this study, and for setting up such a fair test of this hypothesis.

\section*{References}

\refer Arp, H., Bi, H. G., Chu, Y., Zhu, X., 1990, A\&A 239, 33

\refer Broadhurst, T. J., Ellis, R. S., Koo, D. C., Szalay, A. S.,
1990, Nature 343, 726

\refer Burbidge, E. M. \& Burbidge, G. R., 1967, ApJ 148, L107

\refer Burbidge, G. \& Napier, W. M., 2001, AJ 121, 21

\refer Colless, M. \etal, 2001, MNRAS 328, 1039

\refer Croom, S. M. \etal, 2001, MNRAS 322, L29

\refer Efron, B., 1979, Ann.\ Stat.\ 7, 1

\refer Karlsson, K. G., 1990, A\&A 239, 50

\refer Lewis, I.J, \etal, 2002, MNRAS 333, 279

\end{document}